\documentclass[aps,prb,twocolumn,showpacs,preprintnumbers,amsmath,amssymb,superscriptaddress,floatfix]{revtex4}
\usepackage{graphics,bm}
\usepackage{amssymb}
\usepackage{epsfig}
\usepackage{epsf}
\begin{document}

\title{ARPES, Neutrons, and the High-$T_c$ Mechanism}

\author{Wei-Cheng Lee}
\email{leewc@ph.utexas.edu}

\author{A.H. MacDonald}
\email{macd@ph.utexas.edu}

\affiliation{Department of Physics, The University of Texas at Austin, Austin, TX 78712, USA}

\date{\today}

\begin{abstract}
Extensive ARPES and low-energy inelastic neutron scattering studies of cuprate superconductors 
can be successfully described using a weak-coupling theory in which
quasiparticles on a square lattice interact via scalar and spin-dependent effective interactions.
In this article we point out that in underdoped Bi2212 both 
probes are consistent with dominant near-neighbour Heisenberg interactions.
We discuss the implications of this finding for the mechanism of high-$T_c$ superconductivity.
\end{abstract}

\maketitle


\section {Introduction}
It is evident from many experiments\cite{arpesreview,neutronsreview,stmreview}
that the superconductivity of the high-$T_c$ cuprates can be described at low energies and temperatures
by a BCS theory with effective interactions between
square-lattice quasiparticles that lead to short-coherence-length $d$-wave superconductivity.  After many years of study,
the source of this effective interaction has still not been established with certainty.  The d-wave 
property is naturally associated with near-neighbor interactions, but these could be 
spin-independent and attractive ($V$) or Heisenberg-like and antiferromagnetic ($J$).
The effective interactions could be mediated by a lattice or electronic-fluctuation boson,
as in conventional superconductivity, or fall outside of the 
familiar Eliashberg scenario.  The resonance feature in inelastic neutron scattering\cite{neutronsreview}, 
which appears to be generic in cuprates but absent in conventional superconductors, can be explained\cite{demler,paleerpa,tch,leewc}
if interaction parameters are chosen so that the system is close to an antiferromagnetic 
instability, possibly one driven by strong-on site repulsion $U$.  

In this paper we describe an attempt to draw conclusions about
the relative importance of $U$, $V$ and $J$ low-energy effective 
interactions from the numerical arcana of cuprate superconductivity, 
by requiring quantitative consistency between weak-coupling descriptions of 
inelastic neutron scattering resonance (INSR) and ARPES data in Bi$_2$Sr$_2$Ca$_{1-x}$Y$_x$Cu$_2$O$_8$/Bi$_2$Sr$_2$CaCu$_2$O$_{8+\delta}$ (Bi2212).
To describe the INSR we derive weak-coupling expressions for the $(\pi,\pi)$ 
dynamic spin-susceptibility which are exact for any $U-V-J$ model.
The spin-susceptibility appears as a member of a quartet of coupled response 
functions; we find that a four-channel response-function theory is required for 
any non-zero value of $U$, $V$ or $J$.  

We view the use of a weak-coupling theory, in which the fermionic 
excitations are BCS theory Bogoliubov quasiparticles,
as something which is justified at low temperatures by experiment; there is nothing manifestly exotic 
about the cuprate superconducting state apart from its $d$-wave character 
and the large energy scales.  This work is motivated 
by the expectation that a clearer understanding of the character of 
the low-energy effective interactions might hint at its microscopic origin.
Starting from the assumption that only the $U$, $V$ and $J$ effective interactions are 
important and using ARPES experimental values for the square lattice
energy band which crosses the Fermi energy, we conclude from ARPES antinodal gap values we conclude
that in the moderately underdoped regime
$3J/2 - 2V \sim 250 {\rm meV}$.  Similarly from the ocurrence of the INSR phenomenon we conclude that 
$2 J + U \sim 350 {\rm meV}$.  The proximity of these two energy scales strongly suggests that
the Heisenberg effective interaction $J$ is dominant.
We argue that this finding suggests that superconductivity is  
mediated by short-range antiferromagnetic superexchange interactions between low-energy 
quasiparticles which are a remnant of the parent antiferromagnetic 
Mott-insulator and discuss some of the challenges 
which stand in the way of a completely satisfactory microscopic 
theory of superconductivity in doped Mott insulators.   

\section{Phenomenological Model for the Superconducting State at T=0}

\subsection{Effective Hamiltonian}
We consider a low-energy effective Hamiltonian for underdoped Bi2212,
\begin{equation}
H=\sum_{\vec{k}}\epsilon(\vec{k})-\mu\,+\,H_U\,+\,H_V\,+\,H_J
\label{model}
\end{equation}
where $\epsilon(\vec{k})$ is the band energy and the interaction terms are 
\begin{equation}
\begin{array}{l}
\displaystyle
H_U=U \sum_{i} \hat{n}_{i\uparrow}\hat{n}_{i\downarrow}\,\,\,,\,\,\,H_V=V \sum_{<i,j>\sigma\sigma'}\hat{n}_{i\sigma}
\hat{n}_{j\sigma'},\\[2mm]
\displaystyle
H_J=J \sum_{<i,j>} \vec{S}_i\cdot\vec{S}_j.
\end{array}
\end{equation}
The angle bracket notation $<i,j>$ is used to specify that the $V$ and $J$ interactions in our 
model are restricted to near neighbours. 
We do not view this phenomenological Hamiltonian as microscopic,
but as what remains after incoherent higher energy electronic fluctuations are integrated out.  
In view of the Luttinger theorem, the chemical potential $\mu$ is nevertheless fixed by
the doping concentration $x=1-\sum_{\vec{k},\sigma} \langle c^\dagger_{\vec{k}\sigma} c_{\vec{k}\sigma}\rangle$.
The effective interaction parameters $U$, $V$, and $J$ are assummed to be at least weakly doping dependent.

\subsection {Antinodal Gap}
The order parameter for the $d$-wave superconducting state is $\langle c_{i\uparrow}c_{i+\hat{\tau}\downarrow}\rangle=(-)^\tau \Delta$,
where $(-)^\tau= +1$ for $\hat{\tau}=\pm \hat{x}$ and $-1$ for $\hat{\tau}=\pm \hat{y}$, accounting for the $d$-wave character. 
Applying BCS mean-field theory to Eq.(~\ref{model}) leads to a gap function $\Delta(\vec{k})= V_p\Delta\left(\cos k_x-\cos k_y\right)$ 
and to band energies $\epsilon \to \epsilon'$ which are modified by interactions.  Here
\begin{equation} 
V_p=3J/2 - 2V
\end{equation}
has contributions from both of the interactions which can induce $d$-wave superconductivity.
Mean-field ground state properties are completely determined 
by $\epsilon'(\vec{k})$ and $V_p$.
The BCS Hamiltonian yields quasiparticles energies
$\pm E(\vec{k})=\pm (\epsilon'^2(\vec{k})+\Delta^2(\vec{k}))^{1/2}$ which are measured in ARPES experiments.
For $\epsilon'(\vec{k})$ we use the experimental Bi2212 normal state quasiparticle band structure\cite{normanband}. 
The pairing potential $V_p$ of the $d$-wave superconductor is fixed by setting the  
mean-field maximum gap $\Delta_0=2\vert V_p\vert\Delta$ equal to the ARPES antinodal gap\cite{tanaka,leews}.  
Table \ref{table:one} summarizes $V_p$ values obtained in this way for 
several underdoped Bi2212 samples.  For concreteness we focus our discussion of numerical consistency between
ARPES and INSR on the case of doping $x=.144$, reserving a discussion of doping dependence to the end of the paper. 
For $x=.144$, $V_{p} \sim 250 {\rm meV}$.  One of the central question of cuprate superconductivity theory 
is whether this pairing is due primarily to an attractive spin-independent effective interaction ($V< 0$)
or primarily due to an antiferromagnetic spin-dependent effective interactions ($J>0$). 

\subsection{Competing Orders}
The conclusions reached in this paper depend critically on using the same weak-coupling Hamiltonian to consistently describe ARPES 
quasiparticle data and the INSR feature in neutron scattering experiments. 
As we explain in more detail below, the emergence of a 
spin resonance well below\cite{he} the particle-hole continuum signals an incipient instability in cuprates, 
almost certainly the instability toward the antiferromagnetic state.  In a weak-coupling 
generalized random-phase approximation (GRPA) theory the energy cost of static antiferromagnetic fluctuations $K^{s}$ is the sum of 
quasiparticle and interaction energy contributions.  The quasiparticle contribution $K^{s}_{qp}$ is a property of the mean-field state
and based on ARPES data we can conclude that its value is $\sim 400 {\rm meV}$.
We find below that $K^{s}=K^{s}_{qp} - 2J - U$ and conclude 
from qualitative and quantitative aspects of the INSR phenomenon that $K^{s} \ll K^{s}_{qp}$; 
more quantitatively a value close to $\sim 50 {\rm meV}$ seems likely.  
It follows that $2J+U \sim 350 {\rm meV}$.  This conclusion is consistent with many experiments 
which hint at a close competition\cite{competingorders} between spin and superconducting order in cuprates.
To explain this assessment more completely, it is necessary to describe the weak-coupling theory of 
spin and superconducting fluctuations in d-wave superconductors in greater detail.  

\begin{table}
\caption{Singlet-pairing potential $V_p$ for several underdoped Bi2212 samples. 
The doping $x$ is extracted from experimental $T_c$ data, assuming the 
parabolic relation proposed by Presland {\it et al.} in Ref.[~\onlinecite{presland}].}
\centering
\begin{tabular}{c c c c}
\hline\hline
$x$ & $T_c$ (K) & $V_p$ (meV)& $\mu$ (meV)\\ [0.5ex]
\hline 
0.144 & 92 & 250  & -116.467\\
0.126 & 85 & 256 & -111.358 \\
0.11 & 75  & 278 & -105.584 \\
0.099 & 65 & 284 & -102.369\\ [1ex]
\hline
\end{tabular}
\label{table:one}
\end{table}

\section {Weak Coupling INSR Theory} 
Because the interactions in our model Hamiltonian are either on-site or nearest-neighbor, 
fluctuations at the high-symmetry wavevector $\vec{q}=\vec{Q}$,
where $\vec{Q}$ is the ordering wavevector of the parent antiferromagnetic Mott insulator,  
can be expressed in terms of a small number of coupled 
channels\cite{demler,tch}. To demonstrate this, we specialize to the $S_z= + 1$ projection of the 
triplet fluctuation spectrum which is the one relevant to the INS measurements. Expanding the Hamiltonian to quadratic
order around the mean-field state leads to the fluctuation Hamiltonian\cite{leewc}:
\begin{equation}
\begin{array}{ll}
\displaystyle
H^{f}(t)=\frac{1}{A}\sum_{\vec{p},\vec{k},\vec{q},\sigma}&I(\vec{k}-\vec{p})\left[\delta\langle c^\dagger_{\vec{p}\sigma}c_{\vec{p}-\vec{q}\bar{\sigma}}\rangle
c^\dagger_{\vec{k}-\vec{q}\bar{\sigma}}c_{\vec{k}\sigma}\right]\\[2mm]
\displaystyle
&+J(\vec{k}-\vec{p})\left[\delta\langle c^\dagger_{\vec{q}-\vec{p}\sigma}c^\dagger_{\vec{p}\sigma}\rangle
c_{\vec{k}\sigma}c_{\vec{q}-\vec{k}\sigma} + h.c.\right],
\end{array}
\label{hflu}
\end{equation}
where 
\begin{equation}
\displaystyle
I(\vec{k}-\vec{p})=[-U+2V(\cos(k_x-p_x)+\cos(k_y-p_y))],
\end{equation}
and 
\begin{equation}
J(\vec{k}-\vec{p})=-V[\cos(k_x-p_x)+\cos(k_y-p_y)].
\end{equation}
The momentum-dependent term which appears in both interaction form factors, $I(\vec{k}-\vec{p})$ and $J(\vec{k}-\vec{p})$, can be 
rewritten as a sum of separable contributions using 
\begin{equation}
\cos(k_x-p_x)+\cos(k_y-p_y)=(s_{\vec{k}}s_{\vec{p}}+d_{\vec{k}}d_{\vec{p}}+ss_{\vec{k}}ss_{\vec{p}}+sd_{\vec{k}}sd_{\vec{p}})/2 ,
\end{equation} 
where
\begin{equation}
\begin{array}{l}
\displaystyle
s_{\vec{k}}=(\cos k_x+\cos k_y)\,\,\,\,\,\,,\,\,\,\,\,\,
d_{\vec{k}}=(\cos k_x-\cos k_y), \\[2mm]
\displaystyle
ss_{\vec{k}}=(\sin k_x+\sin k_y)\,\,\,\,\,\,,\,\,\,\,\,\,
\displaystyle
sd_{\vec{k}}=(\sin k_x-\sin k_y).
\end{array}
\end{equation}
At wavevector $\vec{q}=\vec{Q}$, the channels involving sine functions all vanish and we then can identify seven
operators, some with $s$- or $d$-wave form factors 
($s_{\vec{k}}$ or $d_{\vec{k}}$), whose fluctuations are influenced by interactions: 
\begin{equation}
\begin{array}{ll}
\displaystyle
\hat{A}_1 = \frac{1}{\sqrt{N}} \sum_{\vec{p}} S^+_{\vec{p}}&
\displaystyle
\hat{A}_2= \frac{1}{\sqrt{2N}} \sum_{\vec{p}} s_{\vec{p}} S^+_{\vec{p}}\\[0mm]
\displaystyle
\hat{A}_3 = \frac{1}{2\sqrt{N}} \sum_{\vec{p}}  d_{\vec{p}} \left(D_{\vec{p}}+\bar{D}_{\vec{p}}\right)&
\displaystyle
\hat{A}_4 = \frac{1}{2\sqrt{N}} \sum_{\vec{p}}  d_{\vec{p}} \left(D_{\vec{p}}-\bar{D}_{\vec{p}}\right)\\[0mm]
\displaystyle
\hat{A}_5= \frac{1}{\sqrt{2N}} \sum_{\vec{p}} d_{\vec{p}} S^+_{\vec{p}}&
\displaystyle
\hat{A}_6 = \frac{1}{2\sqrt{N}} \sum_{\vec{p}}  s_{\vec{p}} \left(D_{\vec{p}}+\bar{D}_{\vec{p}}\right)\\[0mm]
\displaystyle
\hat{A}_7 =  \frac{1}{2\sqrt{N}} \sum_{\vec{p}}  s_{\vec{p}} \left(D_{\vec{p}}-\bar{D}_{\vec{p}}\right)&
\end{array}
\label{operators}
\end{equation}
where $S^+_{\vec{p}}= c^\dagger_{\vec{p}\uparrow} c_{\vec{p}+\vec{Q}\downarrow}$ is a
spin-flip operator and $D_{\vec{p}}= c_{\vec{Q}-\vec{p}\downarrow}c_{\vec{p}\downarrow}$ and 
$\bar{D}_{\vec{p}}= c^\dagger_{\vec{Q}-\vec{p}\uparrow}c^\dagger_{\vec{p}\uparrow}$ are 
pair annihilation and creation operators.  It follows that $\hat{A}_3$ induces 
d-wave pair amplitude
and $\hat{A}_4$ d-wave pair-phase applications.  The 
two-particle Greens functions which capture the fluctuations of these operators are:
\begin{equation}
\hat{\chi}_{ab}(\vec{Q},\omega)=-i\int dt e^{i\omega t}\theta(t)\langle[\hat{A}_a(t),\hat{A}^\dagger_b(0)]
\rangle.
\end{equation}
We focus\cite{missingch2} on the s-wave spin and d-wave pair fields ($\hat{A}_{1-4}$),
which decouple from the d-wave spin and s-wave pair fields ($\hat{A}_{5-7}$) and  
are responsible for the INSR.  We emphasize that none of the earlier theoretical 
interpretations of the INSR feature at $\vec{q}=\vec{Q}$ have accounted properly for
four-channel coupling which appears when the weak-coupling calculation is 
executed correctly. 

\begin{figure}
\includegraphics{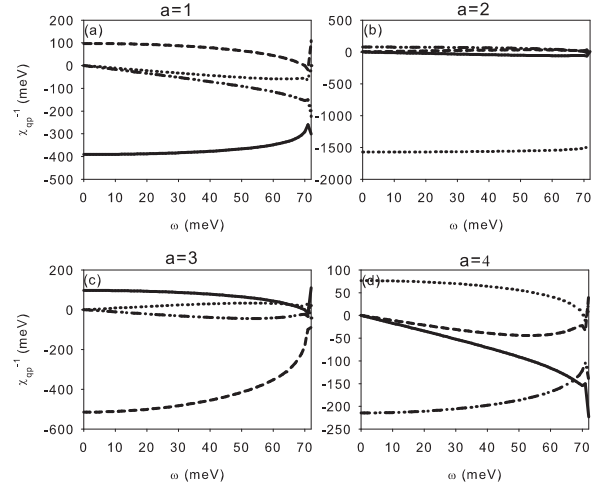}
\caption{\label{fig:one} $\hat{\chi}_{qp,ab}^{-1}(\vec{Q},\omega)$ for $\omega < \Omega_0$ 
where $\Omega_0$ is the gap in the quasiparticle-pair excitation spectrum at $\vec{q}=\vec{Q}$ 
established by $d$-wave order.  For each channel $a$, the solid, dotted, dashed, and dash-dotted
lines represent respectively $\hat{\chi}_{qp,a1}^{-1}, \hat{\chi}_{qp,a2}^{-1},
 \hat{\chi}_{qp,a3}^{-1}$, and $\hat{\chi}_{qp,a4}^{-1}$. 
($\Omega_0 \approx 70 {\rm meV}$ for x=0.144 using the $V_p$ value listed in 
Table \ref{table:one}.)  In a weak-coupling theory these results depend only on 
the normal state band dispersion and on $V_p=3J/2-2V$.  Channel $a=1$ corresponds to spin,
$a=2$ to spin with an extended s-wave form factor, $a=3$ to d-wave pair amplitude and 
$a=4$ to d-wave pair phase.}
\end{figure}

The GRPA Greens functions can be obtained by solving the equation-of-motion for $\hat{\chi}_{ab}(\vec{Q},\omega)$ with the quadratic Hamiltonian given in Eq. \ref{hflu}.  Since BCS mean-field theory with effective interactions describes the low-energy 
charged excitations probed by ARPES, the same theory should also describe the low-energy particle-hole excitations 
probed by inelastic neutron scattering.
We find that 
\begin{equation}
\hat{\chi}^{-1}(\vec{Q},\omega)=\hat{\chi}_{qp}^{-1}(\vec{Q},\omega)-\hat{V}
\label{grpares}
\end{equation}
where $\hat{V}={\rm diag}(-U-2J, J/2-2V,V+J/4,V+J/4)$ 
is the interaction kernel and
\begin{equation}
\begin{array}{l}
\displaystyle
\hat{\chi}_{qp,ab}(\vec{Q},\omega)=\\[2mm]
\displaystyle
\frac{1}{N}\sum_{\vec{k}}\left(\frac{f_a f_b}{\omega-E(\vec{k})-E(\vec{k}')}-
\frac{(-1)^{a+b}f_a f_b}{\omega+E(\vec{k})+E(\vec{k}')}\right)
\end{array}
\label{chi0}
\end{equation}
is the bare mean-field-quasiparticle response function.
In Eq.(~\ref{chi0}) \, $\vec{k}'=\vec{Q}-\vec{k}$, the factor $(-1)^{a+b}$ specifies the simple 
relationship between quasiparticle pair-creation and pair-annihilation matrix-elements\cite{tch} at $\vec{q}=\vec{Q}$,
and the form factors $f_a$ are\cite{tch}:
\begin{equation}
\begin{array}{l}
\displaystyle
f=(p^-(\vec{k},\vec{k}'),\frac{s_{\vec{k}}}{\sqrt{2}}p^+(\vec{k},\vec{k}'),d_{\vec{k}}l^+(\vec{k},\vec{k}'),d_{\vec{k}}l^-(\vec{k},\vec{k}')),\\[4mm]
\displaystyle
p^\pm(\vec{k},\vec{k}')=\frac{\left(u_{\vec{k}} v_{\vec{k}'} \pm v_{\vec{k}} u_{\vec{k}'}\right)}{\sqrt{2}},\\[2mm]
\displaystyle
l^\pm(\vec{k},\vec{k}')=\frac{\left(u_{\vec{k}} u_{\vec{k}'} \pm v_{\vec{k}} v_{\vec{k}'}\right)}{\sqrt{2}}.
\end{array}
\end{equation}
In the GRPA the $\omega$ dependence of $\hat{\chi}^{-1}$ comes entirely from the $\omega$-dependence of $\hat{\chi}_{qp}^{-1}$,
which depends only on the band-structure, on the doping $x$, and on $V_{p}$.  
Typical numerical results for $\hat{\chi}_{qp}^{-1}(\vec{Q})$ are summarized in Fig. [~\ref{fig:one}].

Since the INSR frequency is well below the lower-edge of the 
$\vec{q}=\vec{Q}$ particle-hole continuum, 
it is useful to expand $\hat{\chi}_{qp,ab}(\vec{Q},\omega)$
to leading order in $\omega$:
\begin{equation}
\begin{array}{l}
\displaystyle
\hat{\chi}_{qp,ab}(\vec{Q},\omega)\approx R_0(a,b) - R_1(a,b) \omega+O(\omega^2)\\[2mm]
\displaystyle
R_0(a,b) = \sum_{\vec{k}}\frac{-f_a f_b\left[1+(-1)^{a+b}\right]}{E(\vec{k})+E(\vec{k}')}\\[2mm]
\displaystyle
R_1(a,b) = \sum_{\vec{k}}\frac{f_a f_b\left[1-(-1)^{a+b}\right]}{[E(\vec{k})+E(\vec{k}')]^2}\\[2mm].
\end{array}
\label{r1}
\end{equation}
The leading coupling between even and odd $a$ operators is the Berry-phase coupling
which appears at first order in $\omega$; the most important\cite{demler,tch,leewc} of these is the coupling between 
spin ($a=1$) and d-wave-pair phase ($a=4$).  Even-even and odd-odd fluctuations have no Berry phase 
coupling, but are coupled 
in the static limit. $\chi_{qp}^{-1}$ has a similar low-frequency expansion in which even-even and 
odd-odd fluctuations have relatively little frequency-dependence until $\omega$ approaches the particle-hole continuum closely
as seen in Fig.[~\ref{fig:one}].  The even-even and odd-odd elements of $-\chi^{-1}\equiv K$ specify the energy cost 
of the corresponding particle-hole channel fluctuations while the even-odd elements, approximately linear in frequency, 
specify how the collective fluctuation energy is quantized.  The even-odd elements satisfy $\chi_{qp,ab}^{-1} 
\approx \omega  C_{ab} $. 

\section {Magnetic Plasmon} 
\begin{figure}
\includegraphics{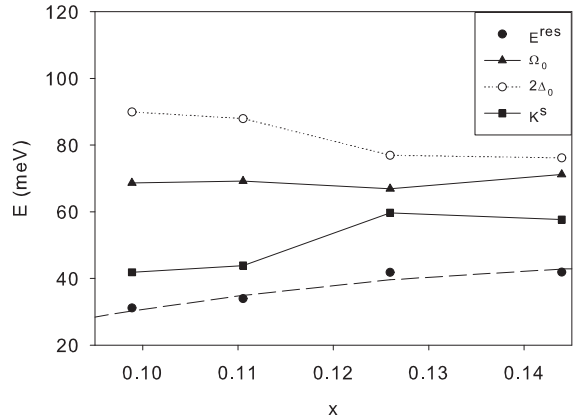}
\caption{\label{fig:three} INSR energy $E^{res}$ calculated from Eq. \ref{detchi} (solid dots) assumming that with $U=V=0$.
When only $J$ is non-zero, its value ($J=\frac{2}{3}V_p$) is fixed by the ARPES antinodal gap.  As shown in this figure 
this assumption predicts the correct value for the INSR position.     
The long-dashed line plots in this figure plot the empirical rule $E^{res}=5.4 k_B T_c$. 
The triangles, white dots, and the black squares show the doping dependencies of $\Omega_0$, and $2\Delta_0$ calculated from $V_p$
and the band structure and the value of $K^s$ obtained when these are combined 
with the ARPES determined value $J$ in the $U=V=0$ model.}
\end{figure}

The INSR energy $E^{res}$ solves 
\begin{equation}
{\rm det}\vert \hat{\chi}^{-1}\vert = {\rm det}\vert \hat{\chi}^{0\,-1}-\hat{V}\vert=0.
\label{detchi}
\end{equation}
The results for $E^{res}$ which are implied by Eq.\ref{detchi} when it is assumed that
the antiferromagnetic $J$ interaction is dominant are summarized in Fig.[~\ref{fig:three}]; the 
same values of $E^{res}$ can be obtained by the time-dependent mean-field theory described in our ealier work\cite{leewc}
which did not specialize to and take advantage of the simplifications possible at $\vec{q}=\vec{Q}$.
The agreement of these results with INSR data is remarkable.  If one interaction 
is assummed to be dominant the model parameters are completely determined by ARPES data. 
{\em The Heisenberg effective interaction model predicts accurate INSR position values over
a broad range of doping.} 

\begin{figure}
\includegraphics{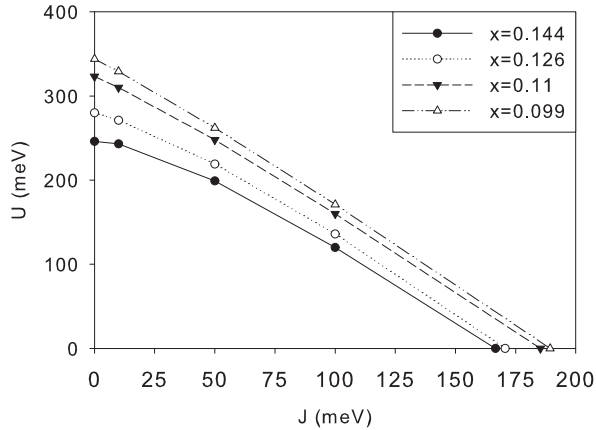}
\caption{\label{fig:two} Lines in $(U,J)$ parameter space which reproduce both the experimental antinodal gaps $\Delta_0$ 
and the INSR energy $E^{res}$ for a series of doping values.  Note that $V = 3J/4-V_p/2$
where the $V_p$ values are listed in Table \ref{table:one}.  For all doping values the intersection of these 
lines with $U=0$ also yields $V=0$.  We argue in the text that this is strongly suggestive that the 
Heisenberg $J$ interaction is dominant.  This figure was constructed using the 4 by 4 expression given in
Eq.~\ref{detchi} for the resonance frequency.}
\end{figure}

To achieve a qualitative understanding of this finding we neglect the ($a=2$) extended-s spin-density fluctuations which 
are much stiffer than other fluctuation modes, as shown in Fig.[~\ref{fig:one}(b)], and 
the weak frequency dependence of the fluctuation energy contributions.  
With these approximations\cite{eresqua}
\begin{equation}
E^{res}\approx \sqrt{\frac{K^s K^{\phi} K^{am} - K^{\phi} (K^{qp}_{13})^2}{K^{am} C^2_{14} + K^{s} C^2_{34} - 2 K^{qp}_{13} C_{14} C_{34}}}
\sim \frac{\sqrt{K^s K^{\phi}}}{C_{14}} 
\label{eplas}
\end{equation}
where $K^s=K^{qp}_{11}-U-2J$, $K^{am}=K^{qp}_{33}+V+J/4$, and $K^{\phi}=K^{qp}_{44}+V+J/4$ are spin, 
$\pi$ amplitude mode, and $\pi$ phase mode stiffnesses respectively.  Since $C_{14} \sim 2$
and the experimental value of 
$E^{res} \sim 40 {\rm meV}$ is small compared to 
$K^{qp}_{11} \approx 400 {\rm meV}$ and $K^{qp}_{44} \approx 200 {\rm meV}$, 
it is clear that interactions must substantially reduce the values of either $K^s$ or $K^{\phi}$, or both. 
In the Heisenberg only model the interaction $J$ reduces $K^s$ to values which are $\sim 50 {\rm meV}$ 
and decrease as expected as the antiferromagnetic state is approached.

The lines in our three dimensional interaction parameter space which are consistent with both 
ARPES and INSR data are illustrated for several doping values in Fig.[~\ref{fig:two}].
We note that the intersection of all lines with $U=0$ also yields $V=0$ to an excellent approximation.
If we imagine that the $U$, $V$ and $J$ effective interactions have independent 
origins, it is natural to expect that one of the three is likely to be dominant.
Our analysis is consistent with this expectation and selects the Heisenberg interaction among the 
three possibilities.  This argument suggests a remarkably simple single-interaction 
low-energy effective model with strength $J$ ranging from $\approx 166$ meV for x=0.144 to $\approx 190$ meV for x=0.099.

The INSR position could of course also be accounted for by fine-tuning both 
$U$ and $V$ at fixed $V_{p}$\cite{unpub}, although we have argued that this is less likely.
For example, if we first assume that d-wave pairing is due entirely to attractive spin-independent effective 
interactions, $V = - V_p/2 \approx - 130 {\rm meV}$.  This value of $V$ results in a small phase stiffness,
$K^{\phi} \sim 70 {\rm meV}$ and would require that $U \sim 300 {\rm meV}$ in order to reduce 
$E^{res}$ into the experimental range.  As we explain later,
this parameter set would correspond to a weaker correlation scenario in which the effective 
value of $U$ is still fairly large.  The INSR {\em magnetic plasmon} in this case has approximately equal 
pair-phase and spin character; in the large $J$ scenario on the other hand the INSR mode has dominant spin 
character because $K^{s} \ll K^{\phi}$.
None of these arguments are sufficently quantitative to 
favor the $V=0$ model choice over the $V=-J/4$ choice, commonly used in 
$t-J$ model calculations and motivated by the theory of the superexchange mechanism. 
Compared to the $J$-only model, this choice shifts $J$ slightly, from  
$J=2 V_p/3$ to $V_p/2$, resulting in slightly larger $K^s$ and smaller $K^{\phi}$
without shifting the INSR position significantly.

We close this section by commenting briefly on the role of inter-channel Berry phase coupling, which has often been 
neglected in theoretical analyses, in determining the INSR position and character.  When only the 
spin-channel is included the equation for the resonance frequency is $K^{s}(E^{res})=0$.  Because of the weak 
energy dependence of $K^{s}_{qp}$ below the particle-hole continuum we see that
when coupling is ignored,
$U+2J$ has to be adjusted to more than $ 90\%$ of $K^{s}_{qp}$ to explain the resonance position, 
placing the system even closer to the antiferromagnetic state instability point.  
For a given value of the interaction stregth, mode coupling substantially lowers the resonance frequency. 
Mode coupling is therefore important in explaining the experimental relationship between the 
value of the resonance frequency and the proximity of the antiferromagnetic state.

\section{Discussion} 

This work starts from the observation that in the low-temperature limit
cuprates act like ordinary superconductors with 
elementary quasiparticle excitations that are describable 
by BCS theory.  ARPES experiments find a Fermi surface with 
area proportional to electron doping $1-x$, rather than hole-doping $x$
and a Fermi velocity which shows no sign\cite{arpesvfvsx} of declining with $x$.
The most remarkable feature of cuprate superconductors at low-temperatures 
is the sharp collective excitation which appears at wavevectors near 
$\vec{Q}=(\pi,\pi)$ in neutron-studies of quantum spin-fluctuations.
The apparently banal character of the low-energy excitations suggests 
that the ground state of these superconductors can be described using 
a weak-coupling theory in which the quasiparticles seen in ARPES experiments 
interact in a way which leads to both d-wave superconductivity and
the inelastic neutron scattering mode.  Given this starting point, it is natural to
ask if quantitative comparison with experiments gives any indication as to the character of the 
presumably renormalized interactions which appear in this low-energy 
effective theory.  In making this assessment, it is important to acknowledge that 
the properties of cuprate superconductors are not entirely universal, and not 
all experimental data is available on any one material.  Nevertheless we 
judge that well established broad trends allow generic conclusions.
The numbers mentioned below are intended to apply most closely to the 
Bi2212 cuprate family. 

In order to make progress toward identifing the renormalized
interactions it seems to be necessary to start by limiting the forms which 
can reasonably be considered.  The fact of singlet d-wave superconductivity on the cuprate square lattice appears to 
require either spin-independent attractive interactions (possibly due to phonons) or spin-dependent
antiferromagnetic Heisenberg interactions (possibly due to short-range antiferromagnetic order)
between electrons on neighbouring sites.  
The proximity of a Mott insulator state for hole-doping $x=0$ would seem to 
require allowing for the possibility of a strongly repulsive on-site Hubbard-like interaction
$U$.  We have therefore considered how the properties of the system depend on 
all three parameters $U$, $V$, and $J$.  The simplest experimental constraint on this
parameter set is imposed by the magnitude of the d-wave antinodal gap which, when 
combined with the band-structure known from ARPES, constrains $V_p= 3J/2-2V$ to a 
value $\sim 250 {\rm meV}$ which increases moderately with decreasing $x$ as summarized in Table I.

An understanding of the implications of the strong INSR at $\vec{Q}=(\pi,\pi)$
requires a more subtle analysis.  In a weak-coupling theory we can determine collective excitation properties by  
examining the influence of interactions on single particle-hole excitations.
As explained in Section III, for the case $\vec{Q}=(\pi,\pi)$ excitations of 
a d-wave superconductor on a square lattice the spin-response function probed 
by neutrons is coupled to pair-magnitude, pair-phase, and extended-s 
spin response functions.  A correct analysis of weak-coupling collective 
fluctuations therefore requires that all four modes be treated simultaneously,
rather than singling out spin-response\cite{paleerpa,rpa1-1,rpa1-2,rpa1-3,rpa1-4,rpa1-5,rpa1-6,rpa1-7} as has been common in RPA theories or 
including only first few modes\cite{demler,tch,rpa3,rpa3-2}.
We emphasize that a theory of the spin-response function requires consideration 
of the four-coupled modes whenever {\em any} of the three interactions we consider is non-zero.
In this theory the energy cost of antiferromagnetic spin fluctuations 
is reduced from $K^{qp}_{11} \sim 400 {\rm meV}$ to $K^s=K^{qp}_{11}-U-2J$ by interactions.
($K^{qp}_{11}$ is purely a property of the BCS mean-field theory and is therefore specified
by the ARPES bands and $V_p$.)  In order for the INSR to appear, as it does in experiment,
below the particle-hole continuum $K^s$ must be reduced to a small value by
interactions, {\em i.e.} the system must be close to an antiferromagnetic instability. 
(A theory which did not accout for coupling between spin and other modes would require that 
$K^s$ be reduced to an even smaller value in order to account for the INSR data.)  On this basis 
we conclude that $U+2J \sim 350 {\rm meV}$.  

Both ARPES and INSR experiments are therefore consistent with a $J$-dominated renormalized interaction model.
The strength for $J$ is comparable to the strength of the superexchange interactions in the undoped antiferromagnetic 
insulator state of the cuprates.  The effective interactions that are consistent with 
experiment are therefore similar to the effective interactions which appear\cite{neutrontheoryreview}
in $t-J$ models, intended to describe quasiparticles which live nearly entirely in 
the Gutzwiller projected\cite{gutzwiller,rmft,rmft2,vanilla,vanilla2} no-double-occupancy Hubbard-model subspace.
The absence of a large Hubbard replusion $U$ in the effective interaction
is also expected for quasiparticles which avoid this strong interaction by 
residing dominantly in the no-double-occupancy subspace.
Any other choice of interaction model requires improbable fine-tuning to achieve 
consistency with the main experimental facts. 

The consistency of the $J$-only interaction model with ARPES and INSR data partially affirms one 
of the main-pictures\cite{neutrontheoryreview} of cuprate superconductivity, namely that it emerges from the antiferromagnetic
interactions in the parent $x=0$ Mott insulator state when holes in the lower Hubbard band become 
itinerant at sufficiently large doping.  There are, however, at this point major challenges which still 
stand in the way of a fully consistent theoretical picture to accompany this scenario.  Mean-field-theory 
slave-boson and other related formulations\cite{neutrontheoryreview,slaveboson} of the 
$t-J$ model, for example, tend to lead to quasiparticle band widths which decreases wtih hole-doping 
and to superfluid density which is strictly proportional to hole-doping $x$.  These predictions are, however, 
clearly at odds with ARPES\cite{arpesvfvsx} and penetration-depth\cite{tempdeppd1,tempdeppd2,tempdeppd3} temperature-dependence
observations, both of which indicate that the Fermi velocity has a weak doping dependence
and that the penetration depth vanishes at a rather large value of hole-doping.
A second theoretical approach motivated by the $t-J$ model
proceeds from the {\em ansatz} that the superconducting state can be described 
by Gutzwiller-projected BCS paired states\cite{rmft,rmft2,vanilla,vanilla2,pbcs,grosanderso}.  
Variational Monte Carlo 
calculations\cite{pbcs} have demonstrated that the 
Gutswiller-projected BCS wavefunction reproduces a variety of quantities observed in experiments.
Furthermore, using the Gutzwiller approximation\cite{gutzwiller},the effect of the Gutzwiller projector
can be replaced approximately by 
doping-dependent band energy and interaction renormalization factors, leading to mean-field equations 
similar to those of slave-boson mean-field theories and of the same form as 
those of the weak-coupling BCS theory.  This simple approximation can 
reproduce the results of the variational Monte Carlo studies to a remarkable degree.
As a result, the {\em plain vanilla} theory helps provides a theoretical explanation for 
weak-coupling low-energy behavior in the superconducting state with effective interaction parameters
that arise in a complex way from the correlated fluctuations of higher energy degrees of freedom.
From this point of view, our main finding in this paper is that the same interactions are 
also quantitatively consistent with the INSR feature.  In addition, as we have shown\cite{leewc} previously,
the pairing wavevector dependence of these low-energy antiferromagnetic fluctuations, provides a 
natural explanation for the fact that the superfluid density goes to zero at a 
finite value of hole-doping.  

One very interesting property of cuprate superconductivities is that the 
Fermi velocity remains nearly constant throughout the underdoped regime, even as\cite{optics} the 
superfluid density and the low-frequency infrared spectral-weight weight decline, 
presumably reflecting in part a decrease in the quasiparticle renormalization factor $Z$.
Although this property is reproduced by Gutswiller projected variational 
Monte-Carlo calculations\cite{pbcs}, it seems surprising from a perturbation 
theory point of view that $Z$ should decline to a small value without a 
corresponding decrease in Fermi velocity; large values for the 
energy-dependence of the self-energy can easily arise from small 
energy denominators associated with resonances near the Fermi energy,
but it is not so obvious how correspondingly
large values for the wavevector-dependence of the self-energy could arise.
We suspect that the quasiparticle renormalization factor in the 
cuprates is not as small as suggested by the Gutzwiller-projected BCS 
state {\em ansatz}.

Another very interesting property of cuprate superconductores is the {\em marginal Fermi liquid}
behavior which often occurs when superconductivity is supressed by temperature (in the overdoped regime) 
or by a magnetic field.  {\em Marginal Fermi liquid} behavior suggests that quasiparticles
are strongly coupled to very-low-energy bosonic excitations, or that the system would be 
very close to a quantum phase transition if superconductivity was suppressed.  
It seems likely to be more than a coincidence that in the low-energy effective model 
we have extracted from experiment, the cuprates
would be very close to antiferromagnetism throughout much of the superconducting dome
if they did not form a superconducting state.
This statement is quantified in the appendix in which we discuss quantitative 
aspects of the competition between superconductivity and antiferromagnetism.  
Decreasing the gap $\Delta_0 \to 0$ decreases $K^{qp}_{11}$ by approximately 
50 meV, just the amount required to reach the antiferromagnetic instability.
Because superconductivity supresses antiferromagnetism, pair-condensation 
must raise a part of the electron-electron interaction energy. 
This property may be part of the explanation for 
the apparent increase\cite{optics,sumrule} in kinetic energy in the 
superconducting state. 

In conclusion, we have performed a weak-coupling analysis of ARPES and INSR experiments in Bi2212,
in an effort to indentify an effective interaction model which is consistent with both experiments. 
We find that the doping dependences of the superconducting gap, and the INSR energy $E^{res}$ can 
be consistently explained by a model with near-neighbor Heisenberg interactions with a strength 
that is consistent with superexchange interactions.  This result suggests that strong short-range 
repulsion and incoherent remnants of the antiferromagnetic insulating parent compound are key to 
high-temperature superconductivity. 

\acknowledgements

This work is supported by the Welch Foundation and by the National Science Foundation under grant 
DMR-0547875. The authors would like to thank W.J.L. Buyers, A.V. Chubukov, B. Keimer, 
P.A. Lee, W.S. Lee, M.R. Norman, and O. Tchernyshov 
for helpful discussions, and to acknowledge the computation resources 
provided by the Texas Advanced Computing Center (TACC) at the University of Texas at Austin.

\appendix 

\section{Competition between $d$-wave superconductivity and antiferromagnetism}

\begin{figure}
\includegraphics{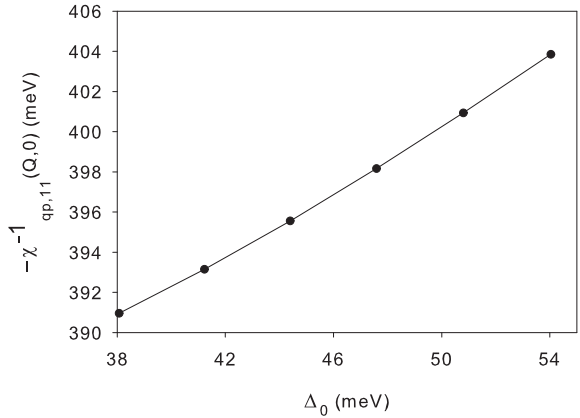}
\caption{\label{fig:four} Numerical evaluation of $-\chi^{-1}_{qp,11}(\vec{k}=\vec{Q},\omega=0)$ as a function of $\Delta_0$ for the doping $x=0.144$ using 
Eq. \ref{chi0}.}
\end{figure}

In this appendix we explain in more detail how $d$-wave superconductivity and antiferromagnetism 
compete in a weak coupling GRPA theory.  As pointed out in an earlier paper\cite{leewc}, 
this competition is responsible for 
a correlation induced suppression of the superfluid density as antiferromagnetism is approached. 

The quantum zero point energy associated with the collective fluctuations probed in INSR experiments is
given approximately by:
\begin{equation}
E^{zp}=\frac{1}{2}\sum_{k}' E^{res}_{\vec{k}}
=\frac{1}{2}\sum_{k}' \frac{\sqrt{K^s_{\vec{k},adia}\,K^{\phi}_{\vec{k},adia}}}{C_{14,\vec{k}}},
\end{equation}
where the prime refers to the sum over wavevectors near $\vec{k}=\vec{Q}$ for which the Berry curvature $C_{14}$ is large.
$K^s_{\vec{k},adia},K^{\phi}_{\vec{k},adia}$ are adiabatic limit stiffnesses which equal $-\chi^{-1}_{qp,11}(\vec{k},\omega=0)-U-2J$ and 
$-\chi^{-1}_{qp,44}(\vec{k},\omega=0)+V+J/4$ respectively. The correlation contribution to the superfluid density is then given by\cite{leewc}:
\begin{equation}
\rho^{cor}= \frac{1}{A}\frac{\partial E^{zp}}{\partial P^2}\approx\frac{1}{4A}\sum_{k}' \frac{E^{res}_{\vec{k}}}{K^s_{\vec{k},adia}}
\frac{\partial K^s_{\vec{k},adia}}{\partial P^2},
\end{equation}
where $P$ is the norm of the pairing wavevector of the BCS wavefunction.  (We explain below why the 
$P$ dependence of $K^{s}$ is much stronger than the $P$ dependence of $K^{\phi}$.) 
Since both $E^{res}_{\vec{k}}$ and  $K^s_{\vec{k},adia}$ are positive quantities, 
$\partial K^s_{\vec{k},adia}/\partial P^2$ determines the 
sign of $\rho^{cor}$. Using the chain rule leads to
\begin{equation}
\frac{\partial K^s_{\vec{k},adia}}{\partial P^2}=\frac{\partial \Delta_0}{\partial P^2}\frac{\partial K^s_{\vec{k},adia}}{\partial \Delta_0}
\end{equation}
Because we know that d-wave superconductivity is weakened
by pair-breaking effect of finite $P$ Doppler shifts on 
nodal quasiparticles we can conclude that $\partial \Delta_0/\partial P^2<0$.

A positive value for $\partial K^s_{\vec{k},adia}/\partial \Delta_0$ characterizes competition between antiferromagnetism and superconductivity. 
When the two orders compete, the energy cost of antiferromagnetic fluctuations is 
increased when superconductivity strengthens. 
In other words, $K^s$ will increase as $\Delta_0$ increases, giving
$\partial K^s_{\vec{k},adia}/\partial \Delta_0>0$ in this case. 
To confirm the competing nature of these interactions, we have performed a numerical calculation of $\Delta_0$ dependence of 
$K^s$. Since $K^s=-\chi^{-1}_{qp,11}(\vec{k},\omega=0)-U-2J$, the $\Delta_0$ dependence of $K^s$ comes entirely from 
$-\chi^{-1}_{qp,11}(\vec{k},\omega=0)$. Fig. [~\ref{fig:four}] plots $-\chi^{-1}_{qp,11}(\vec{k}=\vec{Q},\omega=0)$ as a function of $\Delta_0$ for doping $x=0.144$ as 
an example.  $-\chi^{-1}_{qp,11}(\vec{k}=\vec{Q},\omega=0)$ clearly increases monotonically with $\Delta_0$, consistent with the 
competing order picture discussed above.  It is important to realize that the relative change in $K^s$ is
larger than the 
relative chance in $-\chi^{-1}_{qp,11}(\vec{k}=\vec{Q},\omega=0)$ by apprxoximately a factor of 
ten because of the interaction contributions to the inverse response functions.
We have shown separately\cite{leewc} that the size of this effect combined with the 
reduction in $\Delta_0$ at finite $P$ can explain the large reduction in superfluid densities compared with 
mean-field theory values in underdoped cuprates.   

Our demonstration that superconductivity and antiferromagnetism complete in 
the effective weak-coupling picture of cuprate superconductors contradicts
the the conclusion of Zhang {\it et al.}\cite{zhang} 
who state that weak-coupling theory predicts 
an enhancement of antiferromagnetism due to $d$-wave superconductivity. 
Our calculation shows that their conclusion is not valid when the 
the normal state Fermi surface is similar to what is observed 
in Bi2212 systems. As pointed out by Tchernyshyov {\it et al.}, 
the response functions $\hat{\chi}^0$ at $\vec{q}=\vec{Q}$ are dominated by the {\em hot spots}  
located near $(\pi,0)$.  If this is true, $-\chi^{-1}_{qp,11}(\vec{k},\omega=0) \sim 2\Delta_{antinodal}=2\Delta_0$.
This crude argument is consistent with $\partial K^s_{\vec{k},adia}/\partial \Delta_0 > 0$, 
and very crudely consistent with Fig.[~\ref{fig:four}].  The character of the competition between superconductivity 
and antiferromagnetism is dependent on the normal state band structure.  This dependence is likely responsible 
for some of the differences between hole-doped and electron-doped cuprates.

\end{document}